\documentstyle[11pt]{article}

\input tcilatex

\begin{document}

\title{{\bf New Hadrons Formed by the Fourth SM Family and Iso-singlet Quarks }}
\author{{\bf Hakan C\.{I}FTC\.{I}}$^{\text{a}}${\bf \ and Saleh SULTANSOY}$^{\text{%
a,b}}${\bf \ } \\
$^{a}$ {\it Department of Physics, Faculty of Arts and Science, Gazi
University, }\\
{\it 06500 Teknikokullar, Ankara, TURKEY}\\
$^{b}$ {\it Institute of physics, Academy of sciences, H. Cavid Ave 33, }\\
{\it Baku, AZERBA\.{I}JAN}}
\maketitle

\begin{abstract}
The properties of new heavy mesons containing new heavy quarks have been
investigated. As an example, the fourth SM family quarks and weak
iso-singlet quarks predicted by E$_6$ GUT are considered. Production of
these hadrons at TeV energy lepton colliders have been analyzed.
\end{abstract}

\textwidth 16cm \textheight 21cm \oddsidemargin=1cm \evensidemargin=1cm

\section{Introduction}

It is known that $t$ quark does not form meson and baryon states because of
the high value of $m_t$ and full strength of $tbW$ vertex. On the other
hand, there are strong reasons that the fourth Standard Model (SM) family
should exist [1,2]. We expect that the masses of the fourth family quarks
lie between 300 and 700 GeV with preferable value $m_4=4g_W\eta =8m_W\cong
640$ GeV [3]. New heavy quarks are also predicted by various extensions of
SM and the most wide-known examples are weak iso-singlet quarks predicted by
E$_6$ GUT [4], which is favored by super string theory [5]. In spite of the
fact that the masses of new quarks are larger than $m_t$, they can form
hadrons because of the smallness of mixing between new heavy and ordinary
quarks. Indeed, according to the parametrization of mass matrices given in
[6], mixing between the fourth and third family quarks is predicted to be $%
\left| V_{qU_4}\right| \approx 10^{-3}$. Similar situation is expected for
iso-singlet quarks. The condition for forming new hadrons states is [7]

\begin{equation}
m_Q\leq (100GeV)\left| V_{Qq}\right| ^{-2/3}\text{.}
\end{equation}

New quarks will be copiously produced at the LHC. The observation of the
fourth SM family quarks in ATLAS has been considered in [8, 9]. In this
paper, we discuss the properties of new heavy mesons formed by new quarks.
In section 2, the properties and the systematics of new hadrons have been
investigated. Production of these hadrons at TeV energy lepton colliders
have been considered in section 3. Finally, we give some concluding remarks.

\section{New Heavy Mesons}

According to the Eq. (1), fourth SM family and $E_{6}$ isosinglet quarks
have formed hadron states if their mixing with known quarks is sufficiently
small. In the case of fourth family quarks, parametrization given in [6]
satisfies this condition, whereas new hadrons are not formed in the case of
parametrization given in [1]. Concerning $E_{6}$ isosinglet quarks, we do
not have similar parametrization (in this case one deals with $6\times 6$
mass matrix) and one can make only qualitative estimations. For example, if
the lightest isosinglet quark has the mass $m_{D}\cong 0.5$ TeV, new heavy
hadrons are formed for $\left| V_{qD}\right| <0.09.$

In order to calculate the masses of new mesons we use the logarithmic
potential

\begin{equation}
V(r)=A\log (r)+V_0\text{ ,}
\end{equation}
where $r$ is the radial distance between quarks. For two particles system,
Schrodinger equation has the form ($\hbar =c=1$)

\begin{equation}
\left[ \frac{p^2}{2\mu }+V(r)\right] \Psi (r)=E\Psi (r)\text{ ,}
\end{equation}
where $\mu =m_1m_2/(m_1+m_2)$. One can find the binding energy of meson for
arbitrary $n,l$ states as

\begin{equation}
E_{nl}\cong V_{0}+A\eta _{nl}\text{ ,}
\end{equation}
where $\eta _{nl}$ is solution of the equation

\begin{equation}
\left[ -\frac{d^{2}}{dr^{2}}+\log (r)+\frac{l(l+1)}{r^{2}}\right] g(r)=\eta
g(r)\text{.}
\end{equation}
Below, instead of numerical calculations, we use simplified analytical
procedure (for details see [10]) to solve Eq. (5), which gives the results
coinciding with numerical ones with precision of order $10^{-3}$. Thus the
bound state mass of the $q_{i}\overline{q}_{j}$ system is

\begin{equation}
M(q_i\overline{q}_j)_{nl}=m_{q_i}+m_{q_j}+V_0+A\eta _{nl}
\end{equation}
The calculations involve the potential parameters of the model $\left(
A,V_0\right) $ and the quark masses $\left( m_u=m_d,m_s,m_c,m_b\right) $.
The potential parameters and the quark masses are obtained from the
experimental masses of $\Psi (1S)$, $\Psi (2S),$ $\Upsilon (1S),\Upsilon
(2S),$ $D^0(1S)$ and $s\overline{s}(1S)$ by fitting Eq. (6): $(A,V_0)=$
(0.7328 GeV, -0.8684 GeV), $m_u=m_d=$0.367 GeV, $m_s=$0.561 GeV, $m_c=$1.6
GeV, $m_b=$4.7816 GeV$.$ In addition to these parameters, the fourth SM
family up-quark and $E_6$ isosinglet quark masses are chosen as 638.6 GeV
[6] and 0.5 TeV, respectively. With these parameters, we calculate the
masses of bound states of $c\overline{c}$ and $b\overline{b}$, which are
given in Table 1. It is seen that our results are in good agreement with
experimental data [11] (for comparison see [12, 13]).

In Table 2 we present the masses of quarkonia and mesons formed by the
fourth SM family up-quark and $E_{6}$ isosinglet quark, respectively. It is
seen that the decay of $3S$ quarkonia states into mesons containing $u$ and $%
d$ quarks is admitted. In the case of $4S$ quarkonia decays into mesons
containing $s$ quark is added. In our opinion, the best mechanism for the
production of heavy mesons formed by $u_{4}$ and $D$ quarks is the resonance
formation of $3S$ and $4S$ quarkonia at lepton colliders with subsequent
decay into corresponding meson-antimeson states.

\section{Production of New Mesons at Future Lepton Colliders}

The spin averaged Breit-Wigner cross section for a spin-J resonance produced
in the collision of the particles of spin $S_{1}$ and $S_{2}$ [11] is

\begin{equation}
\sigma _{BW}(E)=\frac{2J+1}{(2S_1+1)(2S_2+1)}\frac{4\pi }{k^2}\frac{%
B_{in}B_{out}\Gamma _{tot}^2}{4(E-R_R)^2+\Gamma _{tot}^2}
\end{equation}
where $k$ is the $c.m.$ momentum, $E$ is the $c.m.$ energy, and $B_{in}$ and 
$B_{out}$ are the branching fractions of the resonance into the entrance and
exit channels. In our case, at $E=E_R$ Eq. (7) takes the form

\begin{equation}
\sigma ^{res}=\frac{12\pi }{M^2}B_{in}B_{out}
\end{equation}
where $M$ is the mass of corresponding quarkonium.

Since lepton colliders have the certain energy spread, the average cross
section at $\sqrt{s}\approx M$ can be estimated from

\begin{equation}
\sigma ^{ave}\cong \frac{\Gamma _{tot}}{\Delta E_{coll}}\sigma ^{res}\text{.}
\end{equation}
Here, we take into account that $\Gamma _{tot}\ll \Delta E_{coll}$. Indeed, $%
\Delta E_{coll}=O$ (1 GeV) for muon collider [14] and $\Delta E_{coll}=O$
(10 GeV) for CLIC [15]. Concerning $\Gamma _{tot}$, we use Couloumb
potential in order to estimate partial decay widths to $e^{+}e^{-}$ and $%
W^{+}W^{-}$ (which is dominant one among the decays into fundamental SM
particles [16, 17]) taking into account that distance between quarks and
antiquarks is much less than $1$ $f$ due to large value of new quarks
masses. Results are presented in Table 3. Then, decay width of 3S levels
into meson-antimeson states is expected to be $O$ (100 MeV). Multiplying
average cross-sections with integrated luminisity, which are 50 fb$^{-1}$
per year for muon collider [14] and 200 fb$^{-1}$ per year for CLIC [15],
one can easily obtain numbers of new mesons produced via formation and
decays of 3$S$ resonance per working year, namely, $\sim 1000$ at muon
collider and $\sim 400$ at CLIC for mesons containing $u_4$ and first family
quarks. Corresponding numbers for mesons formed by $D$ and first family
quarks are $\sim 340$ and $\sim 140$, respectively.

If these new mesons are sufficiently long-lived (which means very small
mixing of heavy quarks with light quarks [18,19]) we will observe
corresponding traces in the detector. The decay length is given by $\beta
\gamma c\tau $. In our case, $\gamma \cong 1$ and $\beta \cong 6\cdot
10^{-4} $. For mesons containing $u_4$ quarks (here we assume the dominance
of mixing between fourth and third SM families and take into account that $%
m_{u_4}\gg m_{W,Z}$)

\begin{equation}
\tau \approx \frac{8x_W}{\alpha _{em}\left| V_{u_4b}\right| ^2}\times \frac{%
m_W^2}{m_{u_4}^3}
\end{equation}
where $x_W=\sin ^2\theta _W\approx 0.21$. Therefore, $l=6\cdot 10^{-2}$ m$%
/\left| V_{u_4b}\right| ^2$ and taking into account present experimental
resolution $\sim 100\mu m$ we conclude that $\left| V_{u_4b}\right| $ should
be less than 2.5$\cdot $10$^{-9}$. Similar consideration for meson
containing $D$ quarks leads to sin$\varphi =\left| V_{Du}\right| <1.5\cdot
10^{-9}$ (here we assume the dominance of $D-d$ mixing and taking account
flavor changing interactions [5,20]). In our opinion, this scenario is
hardly to be realized.

Let us finish this section with some remarks on mesons containing $D$ quark.
As mentioned above, in this case, flavor changing neutral current appear at
tree level and for $m_{D}-m_{W}\approx m_{D}-m_{Z}\gg m_{Z}-m_{W}$ one has $%
Br(D\rightarrow u+W)\approx 0.6$ and $Br(D\rightarrow u+Z)\approx 0.4$ [20].
Therefore, we expect $Br(D\rightarrow jet+l^{+}l^{-})\approx 0.012$ and $%
Br(D\rightarrow jet+\nu \overline{\nu })\approx 0.072$ for decay modes which
differ isosinglet quark from the fourth SM family quarks. For a most
spectacular cases, namely one of the mesons decays into jet and $l^{+}l^{-}$
pairs ($l=e,\mu $) or jet and $\nu \overline{\nu }$ pairs and the other one
decays into three jets we expect $\sim 60$ events per working year at muon
collider and $\sim 20$ events at CLIC.

\section{Conclusion}

We show that mesons formed by new heavy quarks can be observed at future
lepton colliders due to formation (and corresponding decay) of heavy
quarkonia. The numbers of events are not huge (hundreds events comparing to
tens thousands $Q\overline{Q}$ pairs which will be produced at LHC [8, 9]).
However, pure experimental environment provide an advantage for clarifying
the properties of new hadrons.

\section{References}

\begin{enumerate}
\item  A. Datta and S. Raychaudhuri, Phys. Rev. D 49 (1994) 4762.

\item  A. Celikel, A. K. Ciftci and S. Sultansoy, Phys. Lett. B 342 (1995)
257.

\item  S. Sultansoy, arXiv:hep-ph/0004271 (2000).

\item  F. Gursey, P. Ramond and P. Sikivie, Phys. Lett. B 60 (1976) 177.

\item  J.L. Hewett and T.G. Rizzo, Phys. Reports 183 (1989) 193.

\item  S. Atag et al., Phys. Rev. D 54 (1996) 5745.

\item  I. Bigi et al., Phys. Lett. B 181 (1986) 157.

\item  E. Arik et al., Phys. Rev. D 58 (1998) 117701.

\item  ATLAS Detector and Physics Performance TDR, CERN/LHCC/99-15 (1999).

\item  H. Ciftci, E. Ataser and H. Koru, in preparation.

\item  D.E. Groom et al., Review of Particle Physics, Eur. Phys. J. C 15
(2000).

\item  N. Barik, S. N. Jena and D. P. Rath, Phys. Rev. D 41 (1990) 1568.

\item  R.N. Faustov, V.O. Galkin, A.V. Tatarintsev and A.S. Vshivtsev, Int.
J. Mod. Phys. A 15 (2000) 209.

\item  http://pubweb.bnl.gov/users/bking/www/mucoll/overview.pdf

\item  http://cern.web.cern.ch/CERN/Divisions/PS/CLIC/Report/Appendix.html

\item  V. Barger et al., Phys. Rev. D 35 (1987) 3366.

\item  A. K. Ciftci, R. Ciftci and S. Sultansoy, arXiv:hep-ph/0106222 (2001).

\item  P. H. Frampton and P. Q. Hung, Phys. Rev. D 58 (1998) 057704.

\item  P. H. Frampton, P. Q. Hung and M. Sher, Phys. Rep. 330 (2000) 263.

\item  S. Sultansoy, Turkish J. of Phys. 22 (1998) 575.
\end{enumerate}

\newpage\ 

Table.1. The masses of the $c\overline{c}$ and $b\overline{b}$ bound states
(in MeV).

\smallskip\ 

\begin{tabular}{|r|r|r|r|r|}
\hline
Level & $c\overline{c}$ calculated & $c\overline{c}$ experiment [11] & $b%
\overline{b}$ calculated & $b\overline{b}$ experiment [11] \\ \hline
1S & 3097 & 3096.87$\pm $0.04 & 9461 & 9460.30$\pm $0.26 \\ \hline
2S & 3686 & 3685.96$\pm $0.09 & 10050 & 10023.26$\pm $0.31 \\ \hline
3S & 4010 & 4040$\pm $10 & 10374 & 10355.2$\pm $0.5 \\ \hline
4S & 4234 & 4159$\pm $20 & 10598 & 10580.0$\pm $3.5 \\ \hline
5S & 4405 & 4415$\pm $9 & 10769 & 10865$\pm $8 \\ \hline
1P & 3523 & 3510.51$\pm $0.12 & 9898 & 9892.7$\pm $0.6 \\ \hline
2P & 3888 & - & 10272 & 10268.5$\pm $0.4 \\ \hline
\end{tabular}

\bigskip

Table 2. The masses of the bound states formed by the fourth SM family

$u_4$ and isosinglet $D$ quarks (in GeV).

\smallskip\ 

\begin{tabular}{|l|l|l|l|l|l|l|}
\hline
Level & $u_4\overline{u}_4$ & $u_4\overline{u}$ & $u_4\overline{s}$ & $D%
\overline{D}$ & $D\overline{u}$ & $D\overline{s}$ \\ \hline
1S & 1277.10 & 638.86 & 639.06 & 999.90 & 500.26 & 500.46 \\ \hline
2S & 1277.69 & 639.45 & 639.65 & 1000.49 & 500.85 & 501.05 \\ \hline
3S & 1278.01 & 639.78 & 639.97 & 1000.81 & 501.18 & 501.37 \\ \hline
4S & 1278.23 & 640.00 & 640.20 & 1001.03 & 501.40 & 501.60 \\ \hline
\end{tabular}

\bigskip\ 

Table 3. Partial decay widths of heavy quarkonia to $e^{-}e^{+}$ (in keV)

and $W^{-}W^{+}$ (in MeV). $1$ and $2$ correspond to $u_4\overline{u}_4$ and 
$D\overline{D}$, respectively.

\smallskip\ 

\begin{tabular}{|r|r|r|r|r|r|r|}
\hline
Level & Mass($1$) & $\Gamma _{e^{+}e^{-}}$($1$) & $\Gamma _{W^{+}W^{-}}$($1$)
& Mass($2$) & $\Gamma _{e^{+}e^{-}}$($2$) & $\Gamma _{W^{+}W^{-}}$($2$) \\ 
\hline
1S & 1277.10 & 35 & 400 & 999.90 & 6.6 & 142 \\ \hline
2S & 1277.69 & 4.4 & 50 & 1000.49 & 0.8 & 18 \\ \hline
3S & 1278.01 & 1.3 & 15 & 1000.81 & 0.24 & 5 \\ \hline
4S & 1278.23 & 0.5 & 6.2 & 1001.03 & 0.1 & 2.2 \\ \hline
\end{tabular}

\end{document}